# Ponderomotive Acceleration by Relativistic Waves

C. Lau[1,*], P. C. Yeh[1], O. Luk[1], J. McClenaghan[1], T. Ebisuzaki[2], and T. Tajima[1]

[1] Dept. of Physics and Astronomy, University of California, Irvine, CA, 92697 USA
[2] RIKEN, Wako, Saitama 351-0198, Japan

## Abstract

In the extreme high intensity regime of electromagnetic (EM) waves in plasma, the acceleration process is found to be dominated by the ponderomotive acceleration (PA). While the wakefields driven by the ponderomotive force of the relativistic intensity EM waves are important, they may be overtaken by the PA itself in the extreme high intensity regime when the dimensionless vector potential $a_0$ of the EM waves far exceeds unity. The energy gain by this regime (in 1D) is shown to be (approximately) proportional to $a_0^2$. Before reaching this extreme regime, the coexistence of the PA and the wakefield acceleration (WA) is observed where the wave structures driven by the wakefields show the phenomenon of multiple and folded wave-breakings. Investigated are various signatures of the acceleration processes such as the dependence on the mass ratio for the energy gain as well as the energy spectral features. The relevance to high energy cosmic ray acceleration and to the relativistic laser acceleration is considered.

## I.  Introduction

Extreme high energy cosmic rays ([1] extending beyond $10^{20}$ eV and certainly beyond the GZK cutoff [2,3]) attract the contemporary interest, because these high energy particles continue to arrive and moreover the genesis of them remains a puzzle. The prevailing theoretical mechanism for high energy cosmic rays has remained to be that of E. Fermi [4,5]. However, the Fermi mechanism is encountering severe theoretical limitations, as have been reviewed [6,7] and an alternative has been suggested [8]. In short, the Fermi mechanism has a hard time to make sufficient energies beyond $10^{19}$ eV due to the large synchrotron energy loss by the multiple of magnetic turnarounds in addition to other difficulties reviewed by [8] and [6]. Additional difficulties associated with the Fermi mechanism have been discussed [7,9,10,11].

The alternative scenario is the acceleration by large amplitude electromagnetic waves excited near a black hole of large mass in active galactic nuclei (AGN), which can be created along the emanating jets [8,12].

[*] calvin.lau@uci.edu

In this work [8], Ebisuzaki *et al*. find that in the supermassive black hole (BH) jets, extremely large amplitude EM waves can be impulsively excited by the episodic magnetic instabilities [13] of the accretion disk of the BH. This alternative mechanism departs from the stochastic successive encounters with magnetic shocks in the Fermi mechanism and invokes a large organized systematic energy release in astrophysical objects. Earlier, a similar philosophy has been explored in [14] and [15]. In [14], large amplitude ultrarelativistic (where ultrarelativistic refers to the regime in which the dimensionless parameter $a_0$ of the waves far exceeds unity). Alfven waves have been considered as potential accelerating media of cosmic rays. In [15], a pulsed ultrarelativistic electromagnetic wave packet was considered as a driving mechanism of compact cosmic ray sources. Ebisuzaki and Tajima revisited such philosophy and developed new insight into large energy releasing astrophysical phenomena [8]. These studies are now based on better understood energy release mechanisms in astrophysical objects such as accretion disks of BH, active galactic nuclei [13]. Such an energy release of a large amount in a very short period of time is amenable to the process of wakefield excitations, as considered by [11], [9], and [10]. The advantages of these prompt acceleration mechanisms are to avoid many of the Fermi mechanism's difficulties.

The amplitude of these large amplitude EM waves are so relativistic that their dynamics includes what are beyond the well-known wakefield acceleration (WA) [16]. Ebisuzaki and Tajima [8] identified the most important process is the ponderomotive acceleration. Based on this they built a theory for the extreme high energy cosmic ray acceleration and some observational correlations. It is of crucial scientific interest and necessity, therefore, to further understand this ponderomotive acceleration process in detail, and in particular, to investigate it via self-consistent fully nonlinear particle simulation. In the present paper, we study the one-dimensional dynamics along the jet and the acceleration processes by these extreme relativistic EM waves. In Sec. II, we describe the relativistic phenomenology associated with ultrarelativistic EM dynamics, including both the wakefield acceleration and ponderomotive acceleration based on a fully self-consistent electromagnetic particle-in-cell (PIC) simulation. In Sec. III, our PIC simulation results are extended to explore energy scaling laws. We remark on some phenomena associated with the energy spectra observed in our PIC simulation in Sec. IV. Our conclusions are drawn in Sec. V, in which we also mention relevant astrophysical implications, future laser experiments, and possible future areas of research.

## II. Relativistic dynamics, wakefield vs. ponderomotive acceleration phenomenology

A relativistic EM wave induces a ponderomotive force $(q\vec{v} \times \vec{B})$ which displaces electrons along its path. The combination of the quick electron response to regain quasi-neutrality and the massive ions' relative immobility allows for the formation of the plasma wakefield [16]. As the mass-ratio between the ions and electrons decreases (or equivalently as the normalized vector potential exceeds the mass-ratio of ions to electrons), we note that the wakefield amplitude decreases and the structure completely disappears when the mass-ratio is unity such as for an electron-positron plasma. This is in line with the physical mechanism for the wakefield formation. As the mass-ratio approaches unity, the ion response becomes comparable to the electron response such that the space-charge perturbation no longer exists to generate the wakefield [15]. In this limit, PA becomes the dominant mechanism of acceleration for the charged particles. We shall now describe the simulation approach and its characteristics results in detail.

**Description of simulation**

Simulations have been carried out with a one-dimensional relativistic electromagnetic Particle-in-Cell code [17]. The code has one position component $(x)$ and three velocity components $(v_x, v_y, v_z)$ for particles. Both electric and magnetic fields have three components: $(E_x, E_y, E_z)$ and $(B_x, B_y, B_z)$, which are the functions of the coordinate $x$. We adopt the system unit of $c = 1$, $\omega_p = 1$. Field quantities $(E_x, E_y, E_z, B_x, B_y, B_z)$ of Maxwell's equations are solved using finite difference method. $B_x$ is treated as the constant value in this code. The particle positions are updated according to the relativistic equation of motion using the Buneman-Boris scheme [18,19]. We set periodic boundary conditions for both particles and field quantities. In the initial condition, the momenta of electrons perturbed as $p_y = p_{random} + p_0 \cos(k_x(x - x_0))$, where $p_0 = m_e c a_0$, while the momenta of protons to be zero. Here, $p_{random}$ is the momentum corresponding to the thermal motion. We use a linearly polarized wave as the pump wave, $E_y = E_0(x - x_0) \sin(k_x(x - x_0))$ and $B_z = B_0(x - x_0)\sin(k_x(x - x_0))$, where $E_0(x - x_0)$ and $B_0(x - x_0)$ are the envelope functions:

$$E_0(x - x_0) = \frac{m_e \omega_0 c}{q} a_0 \sin^2(k_e(x - x_0)), \quad \text{for } 0 < k_e(x - x_0) < \pi$$

$$B_0(x - x_0) = E_0(x - x_0) \frac{k_x c}{\omega_L}$$

In order to verify the code, we mode analyze those eigenmodes missing from the thermal noise. In addition, in order to examine the extremely relativistic regime, we carried out a severe test of reproducibility of the exact nonlinear solution of the wakefield by Berezhiani and Murusidze [20]. While their solution is a propagating steady solution, our simulation is a temporary growing solution. Thus, our test shows that the nonlinear wakefield wave profile and the scaling of the amplitude of wakefield as a function of $a_0$ are well reproduced. However, as in the simulation the wave grows from the zero value, the maximum amplitude overshoots and saturated back to a lower amplitude in later times. One important observation in our simulation is that the wakefield profile agrees to the spatial position behind the laser pulse until the wavebreaking sets in. Since the solution [20] does not take into account the multi-valued wavebreaking, once the wavebreaking occurs, the analytical and computational solutions depart radically, as expected. The phenomenon of wavebreaking has been studied since the pioneering work by Dawson [21] and Davidson [22]. More discussions on the wavebreaking will follow.

In Fig.1, we show three frames of the longitudinal momentum gain behind the EM pulse that excites a ponderomotive push of electrons ahead of the laser pulse in all three cases, with the mass ratio of ions to electrons ranging from unity to 1836. When the mass ratio is that of the proton, Fig. 1-(c), in addition to the PA pulse near $x = 120\ [c\ \omega_p^{-1}]$, there are a few WA accelerated undulations of electrons near $x \sim 80\ [c\ \omega_p^{-1}]$ and $x \sim 100\ [c\ \omega_p^{-1}]$. The latter undulations are the signature of the WA and its nonlinear evolution behavior showing multiple wavebreaking structures in the electron phase space. In Fig.1-(b), this competition between the PA and WA are clearly contrasted and the structure of the multiple wavebreaking is nearly visible. However, in Fig.1-(a) with mass-ratio = 1, there is no charge restoration between the positrons and electrons in the longitudinal direction and thus we see only the phase space modulation due to the PA near $x = 120\ [c\ \omega_p^{-1}]$. These multiple wave-breakings (non-single valued functions) in the region of the large wakefield excitation cannot be easily theorized, but are likely one of the factors which damp the maximum kinetic energy gained by particles.

In Fig.2, we now scan the laser intensity through the normalized vector potential $a_0$ to study properties of the PA and WA for a fixed mass ratio (in the shown case, mass-ratio=1836). From Figs. 2-(a) to 2-(c) we increase the parameter $a_0$ from 0.4 (sub-relativistic) to 2 (slightly relativistic) to 60 (highly relativistic). In the sub-relativistic case, we see the textbook WA excitation and structure formation [16,23]. In Fig. 2-(b) while we see a bit of PA at the front of the laser at $x = 60\ [c\ \omega_p^{-1}]$, the acceleration is still dominated by WA. Now we also see the complex phase structures of WA showing multiple wavebreakings. For example, at $x \sim 45\ [c\ \omega_p^{-1}]$, we see at least 4 layers of wavebreakings, rendering highly nonlinear dynamics not easily fit for analysis. In Fig.2-(c), in the highly relativistic regime, we clearly observe the dominance of PA, while WA remains subdominant.

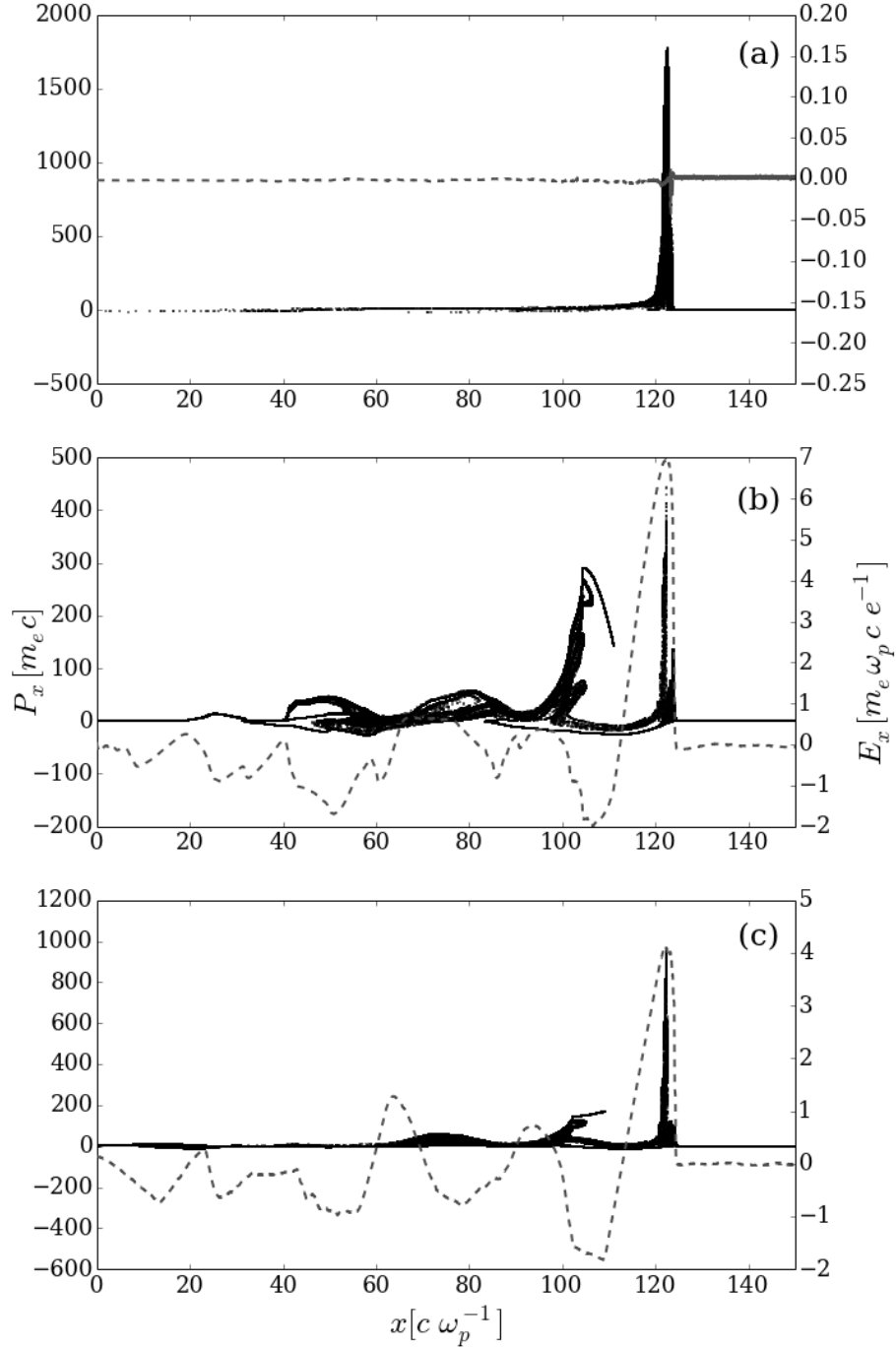

**Figure 1** – The longitudinal $P_x$ momentum gain with relativistic waves, a process dependent on the mass-ratio: phase-space diagrams showing the effect of an electromagnetic wave of same intensity ($a_0 = 20$) on plasmas of differing mass-ratios (a) $\frac{m_i}{m_e} = 1$, (b) $\frac{m_i}{m_e} = 100$, (c) $\frac{m_i}{m_e} = 1836$. The dotted lines represent the electric field in the longitudinal direction (whose amplitude is shown on the right vertical axis, while the momentum gain is shown on the left abscissa). We observe a prominent group of accelerated electrons ahead of the laser pulse due to PA, followed by wakefield accelerated electrons in (b) and (c). In (a), electrons are only accelerated by PA.

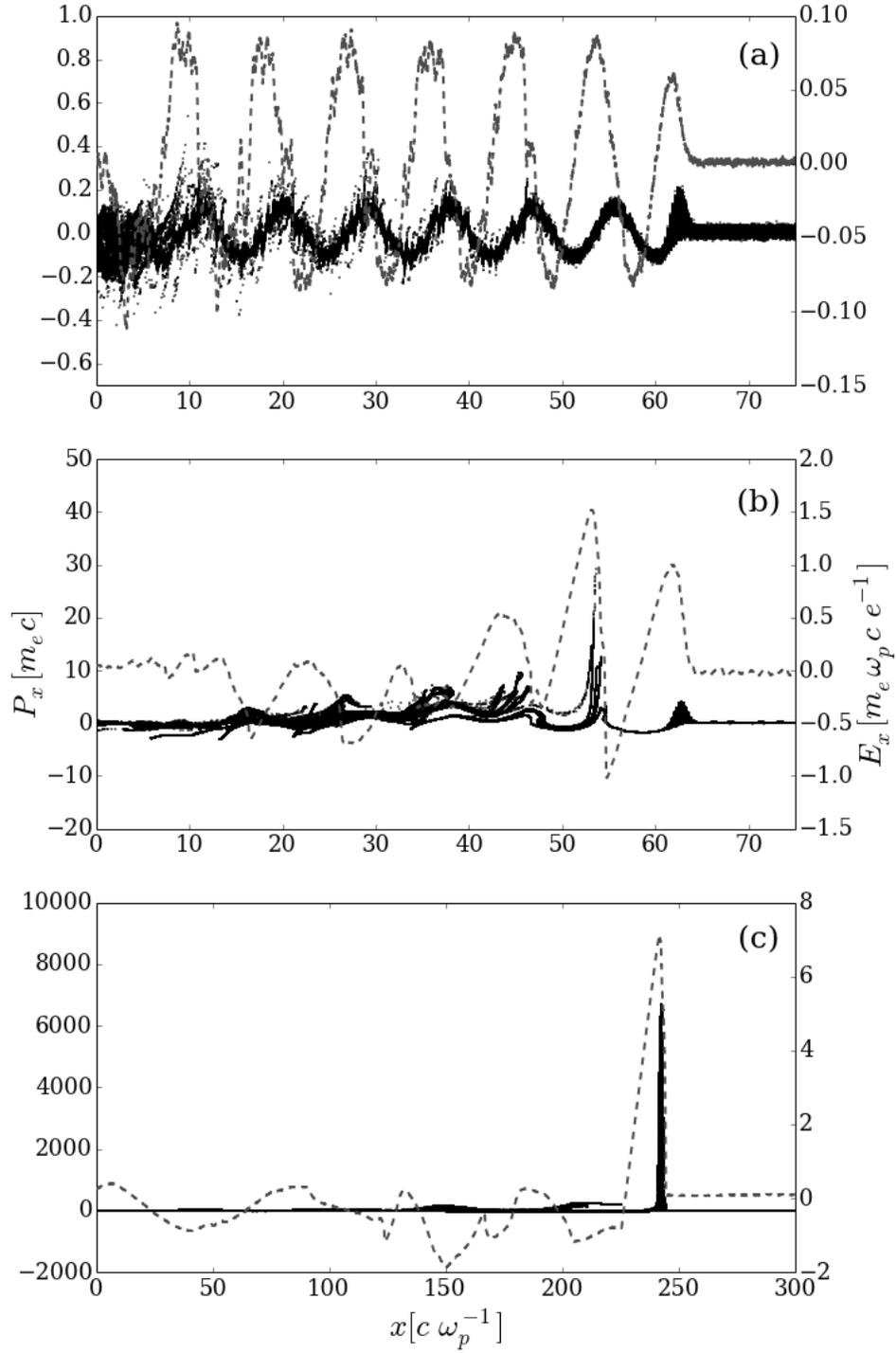

Figure 2 – – The longitudinal $P_x$ momentum gain with relativistic waves, as a function of relativistic wave amplitude: phase-space diagrams showing the effect of electromagnetic waves of increasing intensity on plasmas of same mass-ratios ($\frac{m_i}{m_e} = 1836$), (a) $a_0 = 0.4$, (a) $a_0 = 2$, (a) $a_0 = 60$. The dotted lines represent the electric field in the longitudinal direction. While WA is prominently observed in (a) and (b), PA is predominant in (c).

As the intensity of the EM pulse increases, the behavior of the plasma changes from one dominated by WA [16] to one dominated by PA [8,12,15]. This change in dynamics is due to the change in the ion response. With a low intensity pulse, the more massive ions remain relatively static while the electrons move to respond to the EM pulse. A higher intensity can induce a large enough ponderomotive force such that the ions begin to react with mobility similar to the electrons. When the ion response is comparable, the space-charge density perturbation becomes minimal and this leaves the PA as the main acceleration mechanism of the plasma. That is, as the pulse intensity increases, the dynamics of plasma with high mass-ratio approaches that of plasma with a mass-ratio of unity.

In addition, we also surveyed the density of the plasma by varying the ratio of the EM pulse frequency to the plasma frequency ($\omega_0/\omega_p$). We generally observe an increase of final energy gain with an increased value of $\omega_0/\omega_p$. However, beyond a certain value, the energy gain saturates and even comes back down. We did not extend our work on this beyond these observations for the present work, and for our simulation work in this paper, we kept to a fixed value for $\omega_0/\omega_p$.

### III. Scaling laws of energy gain

Based on our basic parameter surveys of Sec. II, we can summarize the scalings of the ponderomotive force as a function $a_0$, the time to accelerate until the saturated energy $\tau_{W_{max}}$, and most importantly, the electron energy gain as a function of $a_0$. These studies give us a sketch of the properties of PA in contrast to WA.

In Fig.3, we show that regardless of the mass ratio, the ponderomotive force scales linearly with $a_0$. This has been thoroughly anticipated, as in $a_0 \gg 1$, $F_{pond} \sim \frac{q}{c}(v \times B) \propto a_0$, as $v \to c$.

Shown in Fig. 4 is the scaling of the saturation time of electron acceleration. In general, $\tau_{W_{max}}$ is a steep function of $a_0$. However, after the vigorous energy increase, the energy increases in a stage in which a complex mixture of saturation of energy, lull of acceleration, and re-acceleration, etc. Thus, the saturation time of acceleration is not a sharp cutoff time, which leaves complexity in interpreting this scaling.

Figure 5 shows the scaling of the electron energy gain $W_{max}$ as a function of $a_0$. The scaling is nearly asymptotic to $a_0^2$ as $a_0$ increases. This scaling has been anticipated in the work by Ashour-Abdallah [15] and Ebisuzaki-Tajima [8]. Our current study is shown, however, under more controlled conditions with broader parameters. It is also noted that under 1D wakefield acceleration in certain regimes, the accelerated energy is shown to be proportional to $a_0^2$ under proper conditions [23,24,25,26]. This is in general because the relativistic effects enhance both the amplitude of the accelerating fields and the length of acceleration.

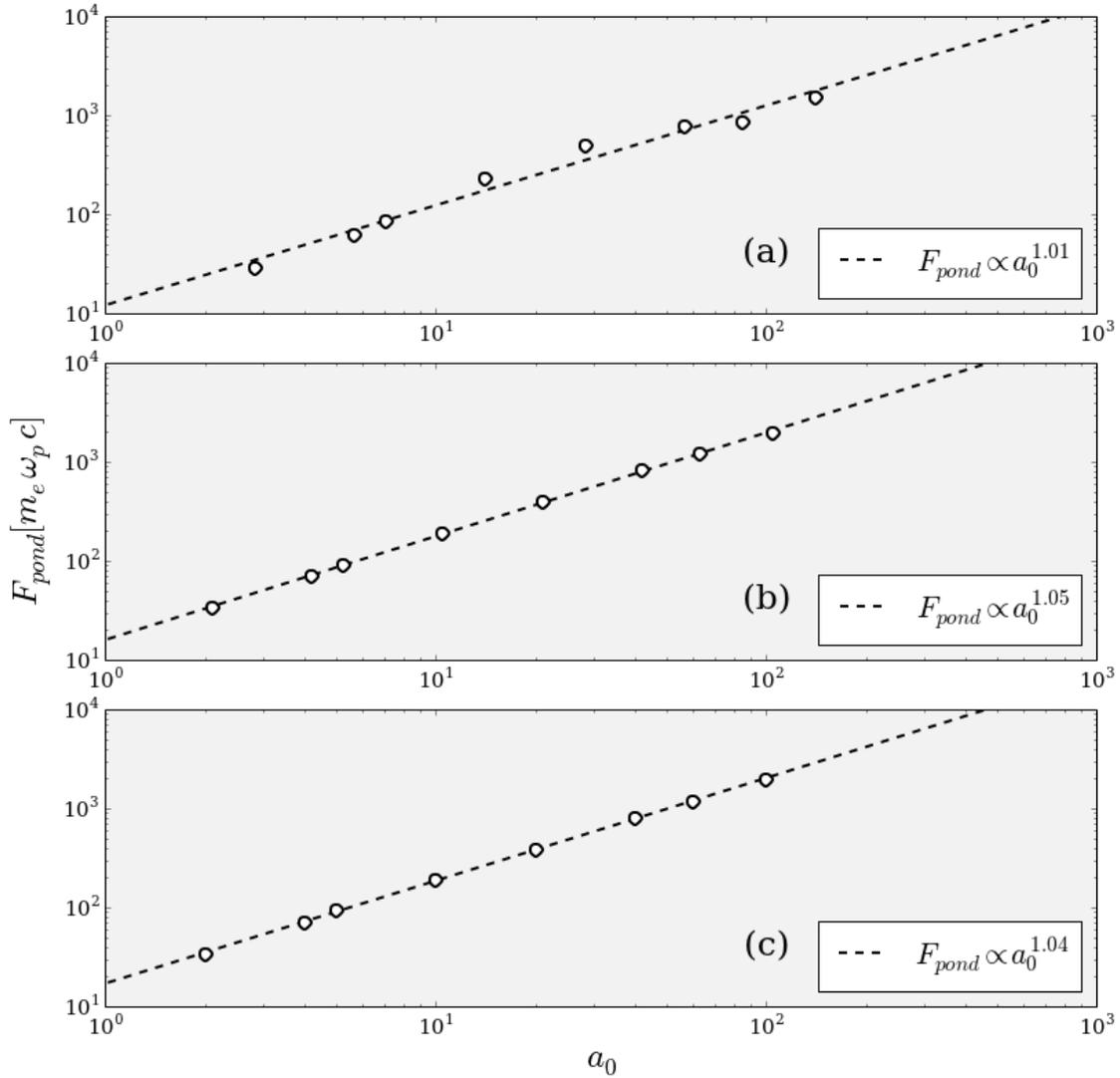

Figure 3- Log-log plots of max ponderomotive force against electromagnetic wave intensity for differing mass-ratios (a) $\frac{m_i}{m_e} = 1$, (b) $\frac{m_i}{m_e} = 100$, (c) $\frac{m_i}{m_e} = 1836$. The ponderomotive force scales linearly with $a_0$, independent of the mass-ratio.

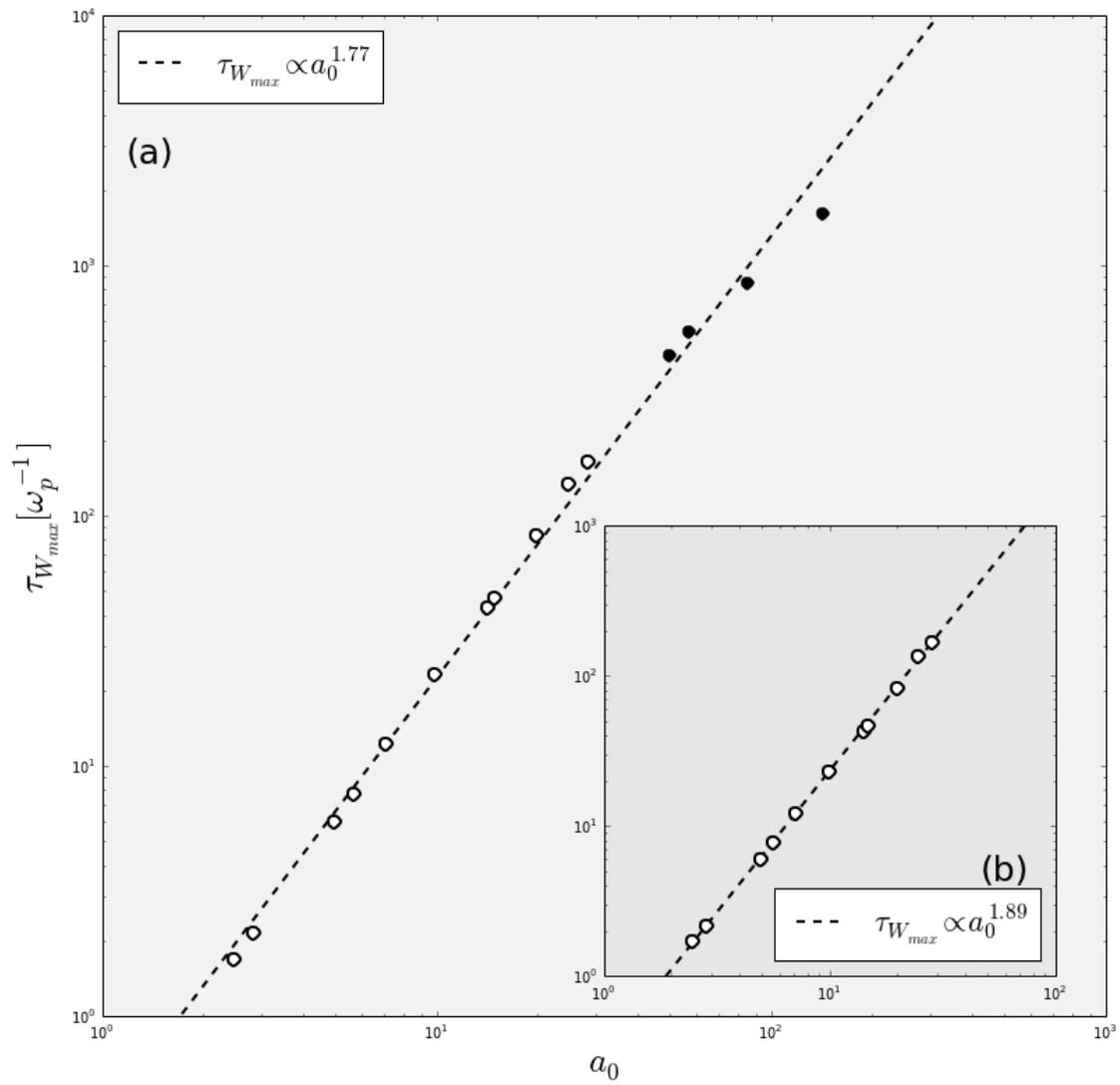

Figure 4 - Log-log plot of time to reach the max kinetic energy against intensity $a_0$ for positron-like mass-ratio. (a) This scaling contains all of the relativistic data points ($a_0 > 1$). (b) This scaling contains points is the region where $a_0$ is relativistic, but finite grid effects have negligible contribution.

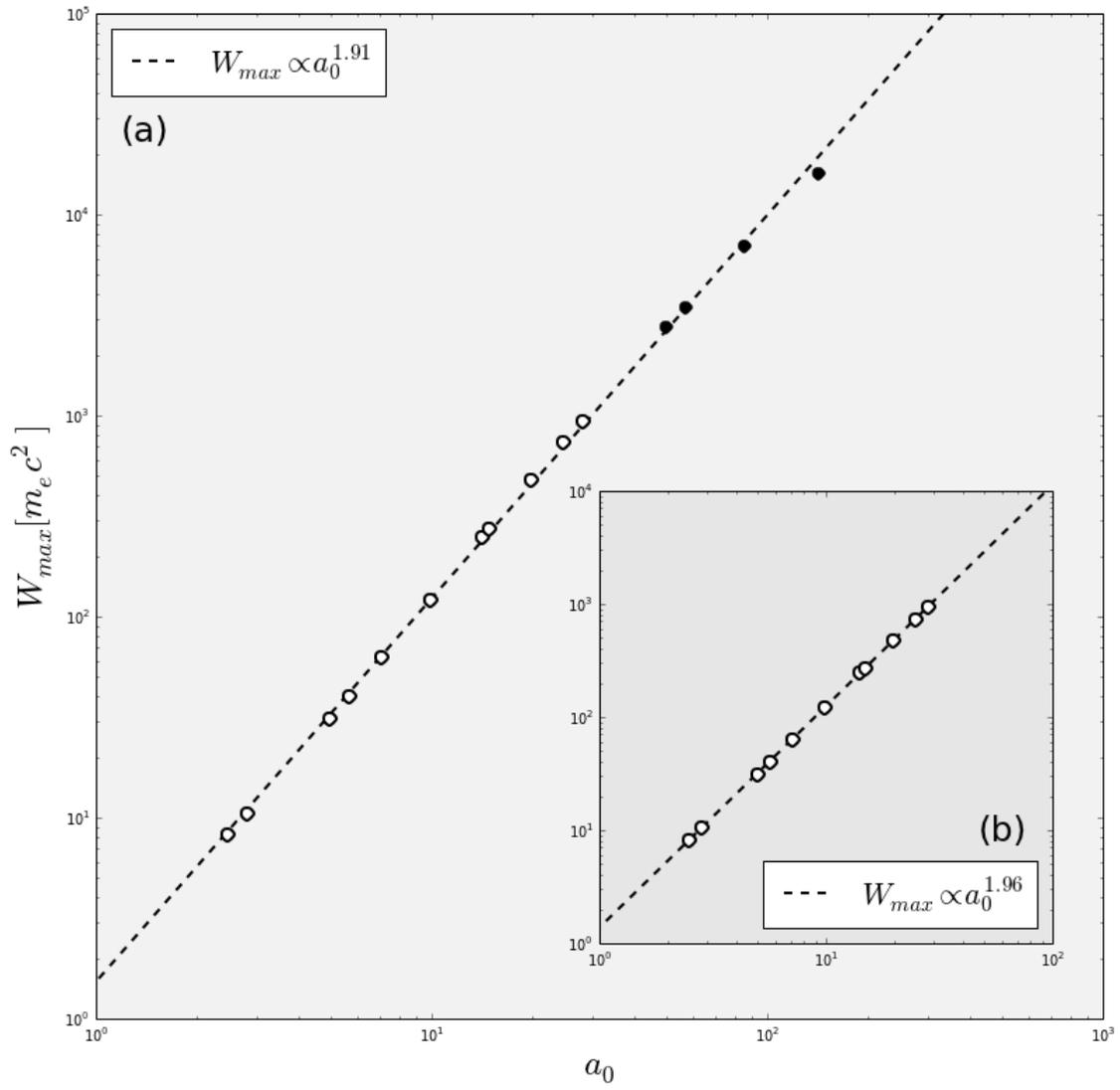

**Figure 5** – The energy gain is studied from the region of the ponderomotive acceleration. Log-log plot of max kinetic energy against amplitude $a_0$ for positron-like mass-ratio. (a) This scaling contains all of the relativistic data points ($a_0 > 1$). (b) This scaling contains points is the region where $a_0$ is relativistic, but finite grid effects have negligible contribution.

## IV. Energy spectra and their implications

In the ultrarelativistic EM regime, the particle acceleration processes of wakefields and ponderomotive fields are highly nonlinear, as exhibited in Sec. II. The observed phenomena include: the coherent pancake density sheet formation (as a manifestation of the relativistic coherence [27]), relativistic nonlinearity [20], the vacuum formation [15], very angular wave profiles [14], multiple wave-folding [28], wave-breaking [21,22], and successive wave-stacking, etc. These nonlinearities not only affect the acceleration process, but also leave imprints on the energy spectra of the accelerated particles. We show some of these characteristics from our PIC studies to illustrate qualitative points.

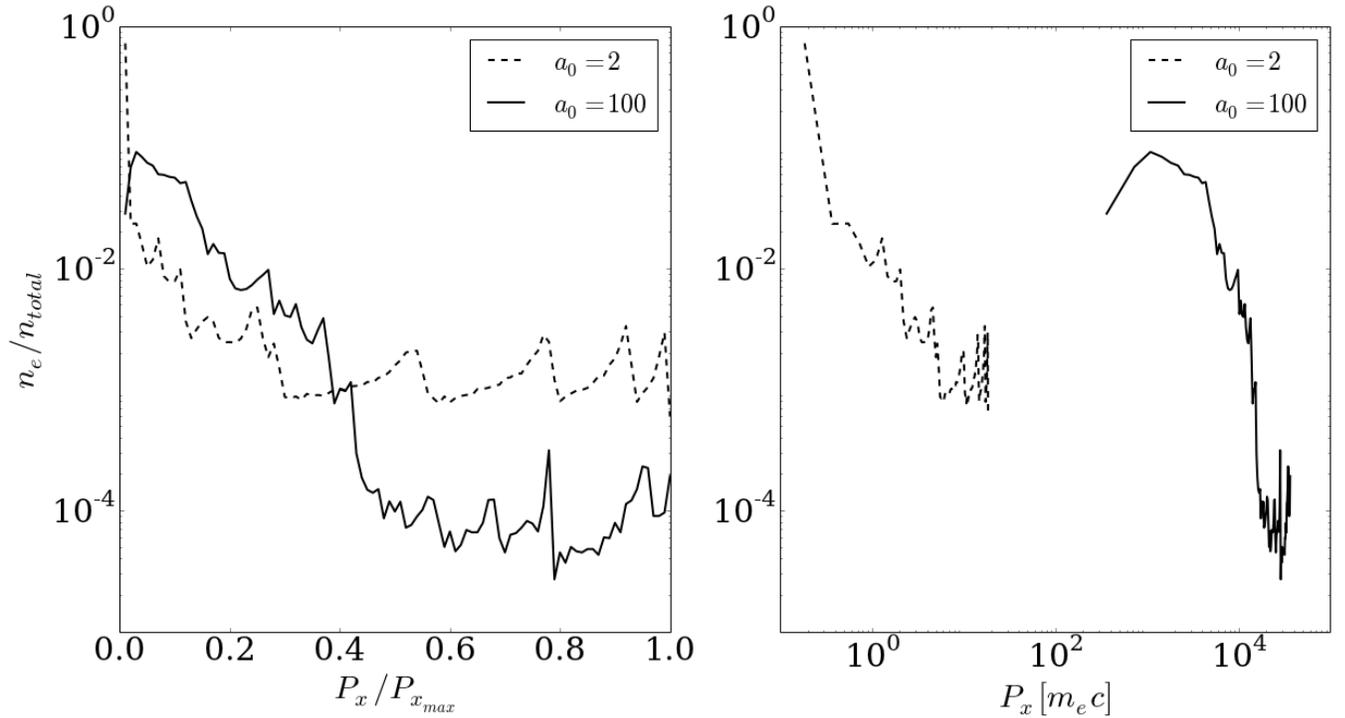

**Figure 6 – The momentum spectra of accelerated particles, ultrarelativistic case with mass-ratio = 1, (a) semi-log, (b) log-log. The extended high energy spectra show peculiar features.**

In Fig.6 and Fig.7, we show the energy spectrum for the two cases of $a_0 = 2$ (mildly relativistic) and $a_0 = 100$ (highly relativistic) for the positron plasma and proton plasma cases, respectively. As discussed, in the mildly relativistic regime, both PA and WA are important in the particle acceleration process.

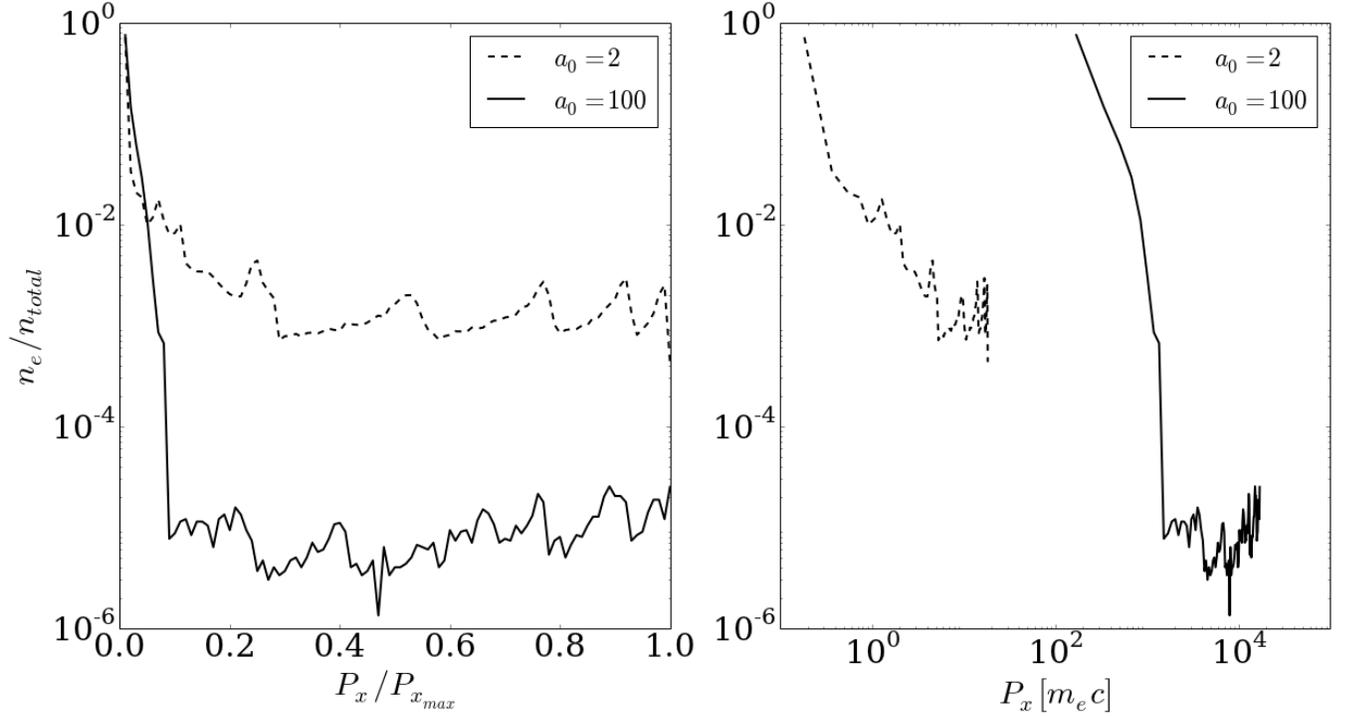

Figure 7 - The momentum spectra of accelerated particles, ultrarelativistic case with mass-ratio = 1836, (a) semi-log, (b) log-log. The high energy spectra are well extended, while they show particular features.

In Fig.6-(a), we see clear bursts of energy (at around $p_x/p_{xmax} \sim 0.5, 0.7, 0.9, 1.0$) for the $a_0 = 2$ case, while for the $a_0 = 100$ case, similar bursts do exist but may be more obscured by other features. In Fig.6-(b), the log-log plot shows the overall energy spectrum. It is noteworthy to see a rather clamped high energy distribution in $a_0 = 100$ case, which is not necessarily a monotonically decreasing function of energy. For the proton mass-ratio case, the analogous plots are shown in Fig.7. Once again, the bursts of energy humps are observed. These humps are believed to be associated with the formation of the multiple-valued wavebreaking features. From the log-log spectrum, we see a significantly large population of high energy components. This is an indication of the efficiency of PA in high energy regimes.

Once we enter the physics of 2D and 3D, we anticipate the overlapping waves with crossing angles. The waves with a crossing angle do not necessarily have a phase velocity close to the speed of light with respect to the predominant wave we are studying (which, in our present case, is the wave in the 1D framework). Thus, this is expected to reduce the predominance of the highest energy components. Furthermore, even in the 1D framework, in astrophysical circumstances, it is more likely to have broader pulse structures than the presently employed "laser" profile. If and when a broader pulse with mixed different phases of such large amplitude Alfven waves (which have now turned into EM waves), it is expected to have acceleration with a shorter dephasing but with incessant repetition of entering in and dephasing with the wave with a different phase. This situation has been studied by Mima et al [29] in 1D. This leads to a power-law spectrum of acceleration with the power index of 2. In higher dimensions, as commented earlier, the tuning-off of highest components is bound to happen, leading to a higher index than 2. It should be noted, however, that the magnetic fields in astrophysical jets and Alfvenic perturbations have a propensity to self-collimate towards the jet axis [13].

In the current study, in order to simplify the physics of the nonlinear PA, we have neglected the radiation effects of electrons and positrons. For example, the betatron oscillations in the wakefield are known to emit copious $\gamma$-rays [30], while these damp the energy of the particles (and reducing the emittance under a certain condition) [31,32]. It is also known [33,34,35] that with a sufficiently high value of $a_0$, the wave emits copious radiation and may not be able to sustain extreme values of $a_0$. At the value of $a_0$ which exceeds $a_0 > \epsilon_{rd}^{-1/3}$ (where $\epsilon_{rd} = \frac{4\pi r_e}{3\lambda}$ with $r_e$ the classical electron radius and $\lambda$ the wavelength of the photon), the wave tends to suffer intense radiation emission [36]. This number may put $a_0$ of intense EM waves less than $\sim 10^8$. A consequence of this may be an efficient mechanism of $\gamma$-ray bursts coming from blackhole jets. In [12], the relativistic EM acceleration from mini-quasars has been studied. In such astrophysical objects, as opposed to supermassive BH, the extremity of the $a_0$ values is much more mitigated and thus the radiative effects are less pronounced. On the other hand, it is also known that the predominantly linearly polarized EM (or Alfven) waves may severely cut radiation emission (as opposed to the case of the circular polarization), turning the threshold $a_0$ higher. We plan to study this issue with far more elaboration in a future study.

## V.    Conclusion

We have demonstrated the robust mechanism of particle acceleration in the regime of very high intensity ($a_0 \gg 1$). In this regime, the predominant acceleration may be borne by the ponderomotive acceleration over the wakefield acceleration. We have elucidated the main properties of this acceleration mechanism. The higher the value of $a_0$ (or equivalently smaller the mass ratio) is, the more predominant role the PA mechanism plays in particle acceleration. In a simplified broad wavefront (ie. 1D), the energy gain of the PA is proportional to $a_0^2$, presenting its vital potential for extreme high energy acceleration.

The emergence of the new laser technique [37] and its application to laser acceleration [38] of recent days may be able to present an experimental realization of the extremely large $a_0$ with an extremely short pulse length, which is favorable for creating the kind of condition we are investigating here. Thus the present PA may be playing an important role to elucidate the physics of both in astrophysical cosmic ray acceleration going beyond the Fermi acceleration and in laboratory laser acceleration to extreme energies.

Because of the expected highly relativistic dynamics (with extreme values of $a_0$), uncharted nonlinear dynamics are bound, including radiative processes. These merit future research. Furthermore, phenomena emerging from astrophysical observation such as polarized $\gamma$-ray emission [39], for example, provide fascinating impetus into the investigations started here that explore the collective nature of particle acceleration in ultrarelativistic regimes through WA and PA. We will also correlate the present process with the temporal bursty characteristics of $\gamma$ emissions believed to be associated with the present extreme high energy acceleration [40]. As remarked, more studies in higher dimensions (2D and 3D) should ensue.

**Acknowledgements**

The authors would like to thank Profs. K. Abazajian, S. Barwick, and G. Yodh on astrophysical discussions, and the authors would also like to thank the High Performance Computing (HPC) Campus Cluster at UC Irvine on which the simulations were performed. This work is supported by the Norman Rostoker Fund. The bulk of the work was carried out as part of the term project of "High Field Science" (PHYSICS 249) in the 2014 winter quarter of UC Irvine.

**Refs.**

1  T.K.Gaisser and T.Stanev, Nucl. Phys.,A. **777**, 98 (2006)

2  K. Greisen, Phys. Rev. Lett. **16**, 748 (1966)

3  G. Zatsepin and V. Kuzmin, JETP Lett. **4**, 78 (1966)

4  E. Fermi, Phys. Rev. **75**, 1169 (1949)

5  E. Fermi, Astrophys. J. **119**, 1 (1954)

6  K. Kotera and A. Olinto, Ann. Rev. Astron. Astrophys. **49**, 119 (2011)

7  A. M. Hills, Ann. Rev. Astron. Astrophys. **22**, 425 (1984)

8  T. Ebisuzaki and T. Tajima, Astropart. Phys. **56**, 9 (2014)

9  P. Chen, T. Tajima, Y. Takahashi, Phys. Rev. Lett. **89**, 161101 (2002)

10  F. Y. Chang, P. Chen, G. L. Lin, R. Noble, R. Sydora, Phys. Rev. Lett. **102**, 111101 (2009)

11  Y. Takahashi, L. W. Hillman, and T. Tajima, in High Field Science, T. Tajima, K. Mima, and H. Baldis, eds., (Kluwer Academic/Plenum, New York , 2000) p.171

12  T. Ebisuzaki and T. Tajima, Euro. Phys. J. **223**, 1113 (2014)

13  T. Tajima and K. Shibata, Plasma Astrophysics (Addison-Wesley, Mass., 1997)

14  J. N. Leboeuf et al., Phys. Rev. A **25**, 1023 (1982)

15  M. Ashour-Abdalla, J. N. Leboeuf, T. Tajima, J. M. Dawson, C. F. Kennel, Phys. Rev. A **23**, 1906 (1981)
16  T. Tajima and J.M. Dawson, Phys. Rev. Lett. **43**, 267 (1979)
17  K. Nagata, Astrophysical Journal, **680**, 627 (2008); http://www-space.eps.s.u-tokyo.ac.jp/~nagata/pic.html,
18  O. Buneman and W. Pardo, Relativ. Plasmas **1**, 205 (1968)
19  J. P. Boris, in Proceedings of the Fourth Conference on Numerical Simulation of Plasmas, Naval Res. Lab, Washington DC, 1970, pp. 3-67.
20  V.I. Berezhiani and I.G. Murusidze, Phys. Lett. **148**, 338 (1990)
21  J. M. Dawson, Phys. Rev. **113**, 383 (1959)
22  R. C. Davidson, Methods in Nonlinear Plasma Theory (Academic, New York, 1972)
23  E. Esarey, C.B. Schroeder, and W.P. Leemans, Rev. Mod. Phys. **81**, 1229 (2009)
24  E. Esarey, A. Ting, P. Sprangle, Phys. Rev. A **42**, 3526 (1990)
25  P. Sprangle, E. Esarey, A. Ting, Phys. Rev. A **41**, 4463 (1990)
26  S. V. Bulanov, V. I. Kirsanov, A. S. Sakharov, JETP Lett **50**, 4 (1989)
27  T. Tajima, Proc. Jpn. Acad. Sci. B **86**, 147 (2010)
28  T. Tajima, D. Habs, and X. Q. Yan, Rev. Accel. Sci. Tech. **2**, 201 (2009)
29  K. Mima, W. Horton, T. Tajima, A. Hasegawa, in Nonlinear Dynamics and Particle Acceleration, Y.H. Ichikawa, and T. Tajima, eds., (American Institute of Physics, New York, 1991) p.27
30  S. Kneip et al., Nature Phys. **6**, 980 (2010)
31  K. Nakajima et al., Phys. Rev. ST Accel. Beams **14**, 091301 (2011)
32  A. Deng, , K. Nakajima, J. Liu, B. Shen, X. Zhang, Y. Yu, W. Li, R. Li, Z. Xu, Phys. Rev. Spec. Top. Accel. Beams **15**, 081303 (2012)
33  G. Mourou, T. Tajima, and S. V. Bulanov, Rev. Mod. Phys. **78**, 309 (2006)
34  S. V. Bulanov, T. Z. Esirkepov, M. Kando, J. K. Koga, S. S. Bulanov, Phys. Rev. E **84**, 056605 (2011)
35  S. V. Bulanov, T. Z. Esirkepov, M. Kando, A. S. Pirozhkov, N. N. Rosanov, Phys. Usp. **56**, 429 (2013)
36  S. V. Bulanov et al., AIP Conf. Proc., 1465, **87** (2012)
37  G. Mourou and T. Tajima, Eur. Phys. J. Spec. Top. **223**, 979 (2014)
38  T. Tajima, Eur. Phys. J. Spec. Top. **223**, 1037 (2014)
39  K. Wiersema et al., Nature **509**, 201 (2014)
40  K. Abazajian, et al. *to be published* (2014)